\documentclass[%
reprint,
superscriptaddress,
showpacs,
preprintnumbers,
amsmath,amssymb,
prc,
floatfix]%
{revtex4-1}

\usepackage{color}
\usepackage{CJK}

\usepackage{graphicx}
\usepackage{dcolumn}
\usepackage{bm}
\usepackage[dvipdfmx,bookmarks=true,colorlinks,%
citecolor=blue,linkcolor=blue,anchorcolor=blue,filecolor=blue,urlcolor=blue,%
           ]{hyperref}

\begin{document}


\begin{CJK*}{UTF8}{gbsn}

\title{Theoretical study of the capture of stable $^{39}$K and 
         neutron-rich \\radioactive $^{46}$K by $^{181}$Ta } 

\author{Bing Wang (王兵)}%
 \affiliation{School of Physics and Engineering, Zhengzhou University, Zhengzhou 
              450001, China}
 \author{Wei-Juan Zhao (赵维娟)}%
 \affiliation{School of Physics and Engineering, Zhengzhou University, Zhengzhou 
              450001, China}
  \author{En-Guang Zhao (赵恩广)}%
 \affiliation{CSA Key Laboratory of Theoretical Physics, 
              Institute of Theoretical Physics, Chinese Academy of Sciences, 
              Beijing 100190, China}
 \affiliation{Center of Theoretical Nuclear Physics, National Laboratory
              of Heavy Ion Accelerator, Lanzhou 730000, China}
\author{Shan-Gui Zhou (周善贵)}%
  \email{sgzhou@itp.ac.cn}
 \affiliation{CSA Key Laboratory of Theoretical Physics, 
              Institute of Theoretical Physics, Chinese Academy of Sciences, 
              Beijing 100190, China}  
 \affiliation{Center of Theoretical Nuclear Physics, National Laboratory
              of Heavy Ion Accelerator, Lanzhou 730000, China} 
 \affiliation{School of Physical Sciences, University of Chinese Academy of Sciences, 
              Beijing 100049, China}
 \affiliation{Synergetic Innovation Center for Quantum Effects and Application, 
              Hunan Normal University, Changsha, 410081, China}

 \date{\today}

\begin{abstract}
The empirical coupled-channel (ECC) model and the universal fusion function (UFF) prescription 
are used to analyse the data of capture cross sections for reactions ${}^{39}$K$+{}^{181}$Ta 
and ${}^{46}$K$+{}^{181}$Ta reported recently by A. Wakhle {\it et al.} 
[\href{https://link.aps.org/doi/10.1103/PhysRevC.97.021602}{Phys. Rev. C 97, 021602(R) (2018)}].
The results of the ECC model are in good agreement with the data of ${}^{39}$K$+{}^{181}$Ta  
while, for ${}^{46}$K$+{}^{181}$Ta, the predictions of the ECC model overestimate the above-barrier 
capture cross sections. Comparing the reduced data of these two reactions, it is found that 
the above-barrier cross sections of ${}^{39}$K$+{}^{181}$Ta are consistent with the UFF 
and are larger than those of ${}^{46}$K$+{}^{181}$Ta. This implies that the capture cross 
sections of ${}^{46}$K$+{}^{181}$Ta are suppressed at energies above the Coulomb barrier. 
Furthermore, at sub-barrier energies, the reduced calculated capture cross sections of 
${}^{39}$K$+{}^{181}$Ta are a little larger than those of ${}^{46}$K$+{}^{181}$Ta, which is 
owing to the coupling to the positive $Q$-value two-neutron transfer channel.

\end{abstract}

\maketitle

\end{CJK*}
 
\section{\label{sec:introduction}Introduction} 
The synthesis of superheavy nuclei (SHN) 
is at the frontier of research in nuclear physics~\cite{Nazarewicz2018_NatPhys14-537}. 
Up to now, superheavy elements with charge number 
$Z \leqslant$ 118 have been produced via
fusion reactions~\cite{Hofmann2000_RMP72-733,Morita2004_JPSJ73-2593,
Oganessian2007_JPG34-165R,Oganessian2010_PRL104-142502}.
However, it is still an open question as to where the center 
of the island of stability is located because the 
SHN produced so far are neutron deficient and still far from the center of the 
predicted island of stability. To synthesize neutron-rich SHN,  
one possible way is to use neutron-rich radioactive beams,
although the intensities of these beams are smaller than 
those of stable beams. In recent years, the synthesis of new heavy nuclei and 
SHN with radioactive beams has been studied a lot~\cite{Loveland2007_PRC76-014612,
Zagrebaev2008_PRC78-034610,Bian2009_NPA829-1,Sargsyan2013_PRC88-054609,
Sargsyan2015_PRC92-054613,Bao2015_PRC91-064612,Bao2016_PRC93-044615}. 

Recently, A. Wakhle {\it et al.} have measured the capture cross sections of 
the reactions ${}^{39}$K$+{}^{181}$Ta and ${}^{46}$K$+{}^{181}$Ta~\cite{Wakhle2018_PRC97-021602R}, 
and the data of these two reactions were compared with the predictions of the time-dependent 
Hartree-Fock (TDHF) calculations and some models including the coupled-channel approach  
of Zagrebaev~\cite{Zagrebaev_ccweb}, the empirical model of Wang and Scheid~\cite{Wang2008_PRC78-014607}, 
and the quantum diffusion approach~\cite{Sargsyan2009_PRC80-034606,Sargsyan2010_EPJA45-125,Sargsyan2011_PRC84-064614}. 
It was found that the calculations of the quantum diffusion approach can do the best overall 
job of representing the capture excitation functions for the reactions ${}^{39}$K$+{}^{181}$Ta 
and ${}^{46}$K$+{}^{181}$Ta, although the calculations of the quantum diffusion approach underestimate 
the sub-barrier capture cross sections of ${}^{39}$K$+{}^{181}$Ta and 
overestimate the above-barrier capture cross sections of ${}^{46}$K$+{}^{181}$Ta.

We have developed an empirical coupled-channel (ECC) model and performed a systematic study of 
capture excitation functions of 217 reaction systems~\cite{Wang2017_ADNDT114-281}. 
In this ECC model, the effects of couplings to inelastic excitations and neutron transfer 
channels are taken effectively into account by introducing an 
empirical barrier weight function~\cite{Wang2016_SciChinaPMA59-642002,Wang2017_ADNDT114-281,Wang2017_NPR34-539_E}.  
The $Q$-value of two-neutron transfer channel for the reaction with stable beam ${}^{39}$K 
is positive while that for the reaction with neutron-rich radioactive beam ${}^{46}$K is negative.
In the present work, we are interested in whether this ECC model can reproduce the data of the  
reactions ${}^{39}$K$+{}^{181}$Ta and ${}^{46}$K$+{}^{181}$Ta. In addition, 
to investigate the effect of the neutron-rich radioactive ${}^{46}$K relative to the stable ${}^{39}$K projectile, 
the data of these two reactions will be reduced and compared with each other through the 
reduction procedure of the universal fusion function (UFF) prescription.
 
The paper is organized as follows. In Sec.~\ref{sec:methods}, we briefly introduce
the ECC model. In Sec.~\ref{sec:results}, the ECC model and the UFF prescription are applied to analyze the 
capture excitation functions of ${}^{39}$K$+{}^{181}$Ta and ${}^{46}$K$+{}^{181}$Ta.  
Finally, a summary is given in Sec.~\ref{sec:summary}.
 
\section{\label{sec:methods}Method} 
The evaporation residue (EvR) cross section for producing 
heavy nuclei via fusion reactions can be written as~\cite{Antonenko1993_PLB319-425,
Antonenko1995_PRC51-2635,Cherepanov1995_NPA583-165,Adamian1997_NPA627-361}
\begin{align}\label{eq:sig_evr}
\sigma_{\rm EvR}(E_{\rm c.m.})=&
                     \sum_{J}\sigma_{\rm capture}(E_{\rm c.m.},J)
                             P_{\rm CN}(E_{\rm c.m.},J) \nonumber\\ 
                             & \hspace*{2mm}\times W_{\rm sur}(E_{\rm c.m.},J),
\end{align}
where $\sigma_{\rm capture}$ is the capture cross section for the
transition of the colliding nuclei over the entrance channel
Coulomb barrier, $P_{\rm CN}$ is the probability of the formation of a compound nucleus (CN) 
after the capture, and $ W_{\rm sur}$ is the survival probability of
the excited CN. It is very important to examine carefully
these three steps in the study of the synthesis mechanism of
heavy nuclei~\cite{Wang2017_ADNDT114-281}. Especially, when heavy nuclei are produced with
radioactive ion beams, one should first examine whether the capture cross section can be 
described well by various models. 

Theoretically, the capture process is treated as a barrier penetration problem. 
The capture cross section at a given center-of-mass energy $E_{\rm c.m.}$
can be written as the sum of the cross section for each partial wave $J$,
\begin{equation}\label{eq:sig_cap}
\sigma_{\rm capture}(E_{\rm c.m.})=\pi\lambdabar^2 
                             \sum_{J=0}^{J_{\rm max}}(2J+1)T(E_{\rm c.m.},J).
\end{equation}
Here $\lambdabar^2= \hbar^2/(2\mu E_\mathrm{c.m.})$ is the reduced de Broglie 
wavelength, $\mu$~is the reduced mass of the reaction system.  
$J_{\rm max}$ is the critical angular momentum. 
$T$ denotes the penetration probability of the Coulomb barrier. 

Comparing with predictions of single barrier penetration model (SBPM), 
sub-barrier capture cross sections are enhanced~\cite{Stokstad1980_PRC21-2427}. 
The enhancement is caused by the strong coupling between the relative
motion and intrinsic degrees of freedom and the coupling to nucleon transfer
channels~\cite{Dasso1983_NPA405-381,Dasso1983_NPA407-221,Broglia1983_PRC27-2433R}.
The capture cross sections can be calculated by either the quantum coupled-channel 
models~\cite{Thompson1988_CPR7-167,Hagino1999_CPC123-143}   
or the empirical coupled-channel (ECC) models. In ECC models, the coupled-channel effects 
are treated effectively by introducing an empirical barrier weight function~\cite{Siwek-Wilczynska2002_APPB33-451,
Zagrebaev2001_PRC65-014607,Liu2006_NPA768-80,Wang2008_PRC78-014607,
Zhu2014_PRC90-014612,Wen2017_CPL34-042501,Bao2017_PRC96-024610,
Wang2016_SciChinaPMA59-642002,Wang2017_ADNDT114-281,Wang2017_NPR34-539_E}.
Besides coupled-channel approaches, the capture can also be described 
by microscopic dynamics models, such as the TDHF theory~\cite{Umar2006_PRC74-021601R,
Umar2012_PRC85-017602,Oberacker2013_PRC87-034611,Dai2014_SciChinaPMA57-1618,
Dai2014_PRC90-044609,Jiang2014_PRC90-064618,Bourgin2016_PRC93-034604}
and the quantum molecular dynamics (QMD) model~\cite{Aichelin1991_PR202-233,
Wang2002_PRC65-064608,Wang2004_PRC69-034608,Feng2008_NPA802-91,
Wen2013_PRL111-012501,Zhu2013_NPA915-90,Wen2014_PRC90-054613,Wang2014_PRC90-054610,
Zanganeh2017_PRC95-034620,Yao2017_PRC95-014607}. As mentioned above, 
the quantum diffusion approach was developed to study the capture process 
as well~\cite{Sargsyan2009_PRC80-034606,Sargsyan2010_EPJA45-125,Sargsyan2011_PRC84-064614}. 
 
Within the ECC model, the coupled-channel effects are taken into 
account by introducing an empirical barrier weight function $f(B)$.
When the interaction potential around the Coulomb barrier 
is approximated by an ``inverted'' parabola, $T$ can be calculated 
by the well-known Hill-Wheeler formula \cite{Hill1953_PR089-1102}.
Then the penetration probability $T$ in Eq.~(\ref{eq:sig_cap}) 
is given as ~\cite{Wang2016_SciChinaPMA59-642002,Wang2017_ADNDT114-281,Wang2017_NPR34-539_E} 
\begin{equation}\label{eq:Tran}
     T(E_{\rm c.m.},J) =\int f(B)T_{\rm HW}(E_{\rm c.m.},J,B){\rm d}B ,
\end{equation} 
where $B$ is the barrier height. Note that there is not a proof or mathematical
derivation of Eq.~(\ref{eq:Tran}) based on the coupled Schr\"odinger
equations. 
Furthermore, for light systems at sub-barrier energies and heavy systems at 
deep sub-barrier energies, the parabolic approximation is not appropriate 
due to the omitting of the long tail of the Coulomb potential. Therefore,
in these cases, the Hill-Wheeler formula does not describe properly the 
behavior of capture cross sections.
In the present work, we are dealing with energies around and 
above the barrier, an energy region where the Hill-Wheeler formula
can be applied. For the barrier penetration
with incident energy much lower than the Coulomb barrier, a new barrier 
penetration formula proposed by Li {\it et al.}~\cite{Li2010_IJMPE19-359} 
can be used. 
  
The empirical barrier weight function $f(B)$ is taken to be an asymmetric Gaussian form
\begin{equation}\label{eq:distri}
f(B)=\left\{
      \begin{array}{cc}
       \frac1N\exp\left[-\left(\frac{B-B_{\rm m}}{\varDelta_1}\right)^2\right],
                                                  \quad & B < B_{\rm m}, \\[1em]
       \frac1N\exp\left[-\left(\frac{B-B_{\rm m}}{\varDelta_2}\right)^2\right],
                                                  \quad & B > B_{\rm m}.
      \end{array}
     \right. 
\end{equation}
$f(B)$ satisfies the normalization condition $\int f(B)dB=1$. 
Thus the normalization coefficient $N =\sqrt{\pi}(\varDelta_1+\varDelta_2)/2 $. 
$\varDelta_1$ and $\varDelta_2$ denote the left width and the right 
width of the empirical barrier weight function. The $B_{\rm m}$ denotes the most probable barrier
height, i.e., the peak of the empirical barrier weight function.

In our ECC model~\cite{Wang2016_SciChinaPMA59-642002,Wang2017_ADNDT114-281,Wang2017_NPR34-539_E}, 
the barrier distribution is related to the effects of couplings to low-lying collective 
states and positive $Q$-value neutron transfer (PQNT) channels. Considering 
the dynamical deformations due to the attractive nuclear force and the 
repulsive Coulomb force~\cite{Zagrebaev2003_PRC67-061601R,Wang2012_PRC85-041601R}, 
a two-dimensional potential energy surface (PES) with respect to quadrupole 
deformation of the system and relative distance $R$ can be obtained. 
To take into account the effects of the couplings to low-lying collective 
states, empirical formulas for calculating the parameters of the empirical barrier 
weight function were proposed based on the PES. Then the effect of the 
coupling to the PQNT channels is simulated by broadening the empirical barrier weight function. 
In the present model, only two-neutron transfer channel is considered. When 
the $Q$-value for two-neutron transfer is positive, the widths of the empirical barrier weight function
are calculated as $ \varDelta_i \rightarrow gQ(2n)+\varDelta_i, (i=1,2)$, 
where $Q(2n)$ is the $Q$-value for two-neutron transfer. $g$ is taken as $0.32$ for all 
reactions with positive $Q$-value for two-neutron transfer channel. 
In addition, this ECC model was extended to describe the complete 
fusion cross sections for the reactions involving weakly bound 
nuclei at above-barrier energies~\cite{Wang2014_PRC90-034612,Wang2016_PRC93-014615}. 
More details for the ECC model can be found in Refs.~\cite{Wang2016_PRC93-014615,Wang2017_ADNDT114-281}.

\begin{figure}[htb!]
\centering{
\includegraphics[width=0.85\columnwidth]{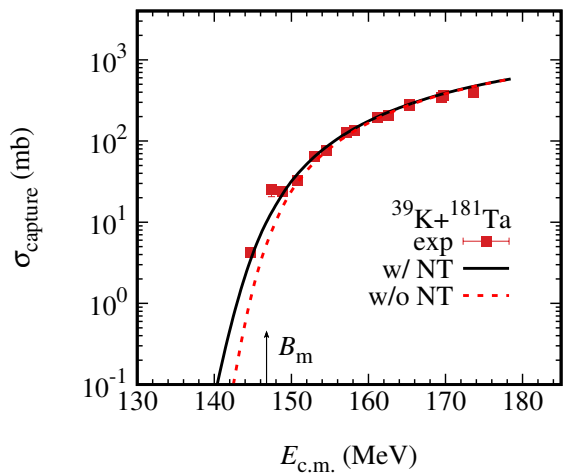}
 }
\caption{(Color online) The calculated and experimental capture cross sections 
of ${}^{39}$K$+{}^{181}$Ta. The dash line denotes the results from the 
ECC calculations without the neutron transfer (NT)
effect considered. The solid line denotes the results from the ECC 
calculations with the NT effect considered. 
The arrow indicates the peak of the empirical barrier weight function $B_{\rm m}$ 
given in Eq.~(\ref{eq:distri}).
The data taken from Ref.~\cite{Wakhle2018_PRC97-021602R} are represented 
by the solid squares.}\label{fig:39K}
\end{figure}

\section{\label{sec:results}Results and Discussions}
Note that the parameters of the empirical barrier weight function are 
calculated by the empirical formulas which were proposed in 
Ref.~\cite{Wang2017_ADNDT114-281} where the parameters of the deformed 
nuclear potential and the Coulomb potential were also fixed. 
Therefore, there is no free parameters in the following calculations.

We first focus on the reaction with stable beam ${}^{39}$K.
The comparison of the calculated capture cross sections to the experimental 
values for ${}^{39}$K$+{}^{181}$Ta is shown in Fig.~\ref{fig:39K}.
The arrow indicates the peak of the empirical barrier weight function $B_{\rm m}$ 
given in Eq.~(\ref{eq:distri}). The solid line denotes the results from the ECC 
calculations with all the couplings (to low-lying collective states and PQNT channels). 
The three parameters $\varDelta_1$, $\varDelta_2$, and $B_{\rm m}$ of the empirical barrier 
weight function are $3.56$~MeV, $23.46$~MeV, and $146.78$~MeV, respectively.
One can see that the results of the ECC model are in good agreement with the data.
The results of this ECC model are much closer to the data than those 
calculations shown in Ref.~\cite{Wakhle2018_PRC97-021602R}.
Note that, for ${}^{39}$K$+{}^{181}$Ta, the $Q$-value of two-neutron transfer channel 
is $3.67$~MeV, thus part of the enhancement of sub-barrier capture cross sections 
comes from the coupling to the PQNT channel. To show this enhancement clearly, 
the results from the ECC calculations without the coupling 
to the neutron transfer channels considered are shown in Fig.~\ref{fig:39K} by the dash line.
Thus, the difference between the solid line and the dash line shows the PQNT effect on capture 
cross sections.

\begin{figure}[tb!]
\centering{
\includegraphics[width=0.85\columnwidth]{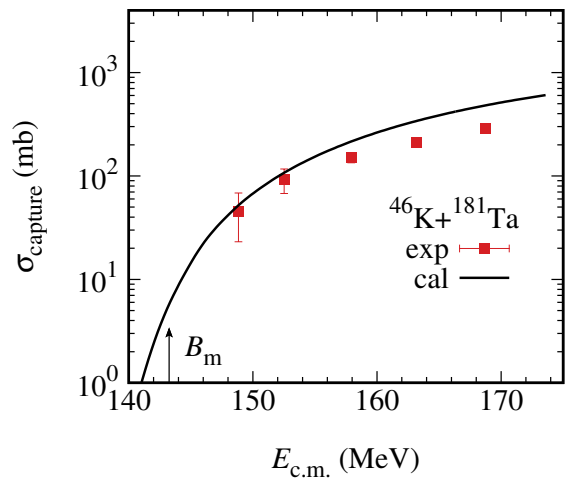}}
\caption{(Color online) The calculated and experimental capture cross sections 
of ${}^{39}$K$+{}^{181}$Ta. The solid line denotes the results from the ECC 
calculations. The arrow indicates the peak of 
the empirical barrier weight function $B_{\rm m}$ given in Eq.~(\ref{eq:distri}).
The data taken from Ref.~\cite{Wakhle2018_PRC97-021602R} are represented 
by the solid squares.}
\label{fig:46K}
\end{figure}

For the reaction with neutron-rich radioactive beam ${}^{46}$K, 
the comparison of the calculated capture cross sections to the experimental 
values is shown in Fig.~\ref{fig:46K}. For ${}^{46}$K$+{}^{181}$Ta, the $Q$-value of two-neutron 
transfer channel is negative, thus, in this case, the coupling to the PQNT channels 
does not affect the capture cross sections. Therefore, only the couplings to low-lying collective 
states is responsible for the enhancement of the sub-barrier capture cross sections. 
The three parameters, i.e., $\varDelta_1$, $\varDelta_2$, and $B_{\rm m}$, of the barrier 
weight function are $2.38$~MeV, $22.20$~MeV, and $143.25$~MeV, respectively.
The results from the ECC calculations are shown in Fig.~\ref{fig:46K} by the solid line. 
It can be seen that the calculated results overestimate the cross sections except the 
two lower energies or, in other words, the above-barrier capture cross sections are suppressed 
as compared with the ECC calculations. The results from the ECC calculations are similar to those 
from the quantum diffusion approach shown in Ref.~\cite{Wakhle2018_PRC97-021602R}.

The results from the ECC calculations are very interesting, as the data of the 
reaction with stable beam ${}^{39}$K are reproduced quite well while those of 
the reaction with neutron-rich radioactive beam ${}^{46}$K are not. Therefore, 
it is natural to ask what is the effect of the neutron-rich radioactive ${}^{46}$K 
relative to the stable ${}^{39}$K projectile? Actually, in Ref.~\cite{Wakhle2018_PRC97-021602R}, 
the capture cross sections of ${}^{39}$K$+{}^{181}$Ta and ${}^{46}$K$+{}^{181}$Ta 
were reduced by the traditional reduction procedure, i.e., $E_{\rm c.m.}\rightarrow E_{\rm c.m.}/V_{\rm B}$ and 
$\sigma_{\rm capture} \rightarrow \sigma_{\rm capture}/R_{\rm B}^2$. The parameters $V_{\rm B}$ and $R_{\rm B}$ were extracted from
the plot of the cross sections vs. $1/E_{\rm c.m.}$.  
It was found that the 
reduced excitation functions of these two reaction do not show any significant difference. 
In the present work, we adopt another reduction method proposed in 
Refs.~\cite{Canto2009_JPG36-015109,Canto2009_NPA821-51} which can eliminate completely 
the geometrical factors and static effects of the potential between the two nuclei. 
In this case, the capture cross section and the collision energy are
reduced to a dimensionless fusion function $F(x)$ and a dimensionless variable
$x$,
\begin{equation}\label{eq:uff}
F(x) = \frac{2E_{\rm c.m.}}{R_{\rm B}^2\hbar\omega}\sigma_{\rm capture}, \quad
   x = \frac{ E_{\rm c.m.}-V_{\rm B}}{\hbar\omega},
\end{equation}
where $V_{\rm B}$, $\hbar\omega$, and $R_{\rm B}$
denote the height, curvature, and radius of the barrier which are calculated by the double folding and
parameter-free S\~ao Paulo potential (SPP)~\cite{CandidoRibeiro1997_PRL78-3270,Chamon1997_PRL79-5218,
Chamon2002_PRC66-014610}. The barrier parameters calculated by the SPP are shown in Table~\ref{tab:para}.

\begin{figure}[htb!]
\centering{
\includegraphics[width=0.8\columnwidth]{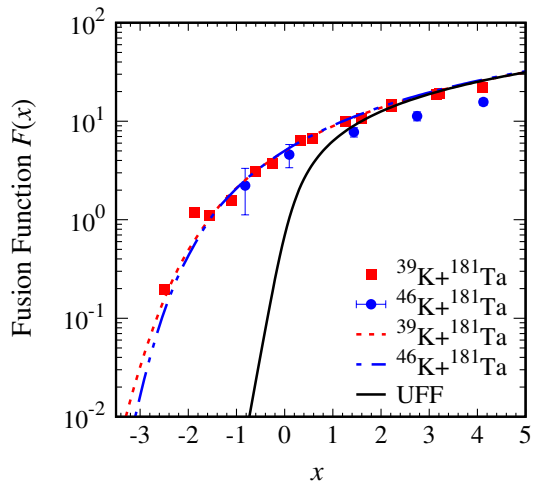}}
\caption{(Color online) The reduced capture excitation function $F(x)$ for reactions 
${}^{39}$K$+{}^{181}$Ta and ${}^{46}$K$+{}^{181}$Ta as a function of $x$.
The dash and dash-dotted lines denote the reduced calculated capture cross 
sections of ${}^{39}$K$+{}^{181}$Ta and ${}^{46}$K$+{}^{181}$Ta, respectively. 
The solid line represents the UFF. The data taken from Ref.~\cite{Wakhle2018_PRC97-021602R} 
are represented by the solid squares and points.}
\label{fig:uff}
\end{figure}
 
The reduced capture excitation functions, i.e., fusion functions $F(x)$, for the reactions 
${}^{39}$K$+{}^{181}$Ta and ${}^{46}$K$+{}^{181}$Ta are shown in Fig.~\ref{fig:uff} 
by the solid squares and points, respectively. It can be seen that, at $x > 0$ region, 
i.e., at above-barrier energies, the reduced capture cross sections
of ${}^{39}$K$+{}^{181}$Ta are clearly larger than those of ${}^{46}$K$+{}^{181}$Ta.
Furthermore, the reduced above-barrier capture cross sections of ${}^{39}$K$+{}^{181}$Ta are 
close to the UFF (denoted by the solid line) 
which are the predictions of the Wong formula~\cite{Wong1973_PRL31-766} reduced by Eq.~(\ref{eq:uff}). 
While the above-barrier capture cross sections of ${}^{46}$K$+{}^{181}$Ta lie below the UFF. 
This tells that the above-barrier capture cross sections of ${}^{46}$K$+{}^{181}$Ta are 
suppressed as compared with those of ${}^{39}$K$+{}^{181}$Ta and the UFF.
This result is not consistent with the conclusion drawn in Ref.~\cite{Wakhle2018_PRC97-021602R}, 
which might result from the different barrier parameters used in the reduction procedures. 
The parameters $V_{\rm B}$ and $R_{\rm B}$ extracted from Ref.~\cite{Wakhle2018_PRC97-021602R} 
are also given in Table~\ref{tab:para}. It can be found that the extracted parameter 
$R_{\rm B}$ of ${}^{46}$K$+{}^{181}$Ta (10.16 fm) is obviously smaller than that 
of ${}^{39}$K$+{}^{181}$Ta (12.82 fm), while from the SPP, the opposite is true, i.e., 
$R_\mathrm{B} = 12.333$ fm for $^{46}$K+$^{181}$Ta 12.030 fm for $^{39}$K+$^{181}$Ta. 
Actually, the barrier parameters extracted from the experimental excitation function already include part of the dynamical effects.
For ${}^{46}$K$+{}^{181}$Ta, the fact that $R_{\rm B}$ extracted from the experiment is samll is a manifestation
of the suppression effect on the above-barrier cross sections. In addition, from Fig.~3 in Ref.~\cite{Wakhle2018_PRC97-021602R}, 
it is shown that the models overestimate the above-barrier cross sections of ${}^{46}$K$+{}^{181}$Ta.
The results shown in Ref.~\cite{Wakhle2018_PRC97-021602R} strongly support the conclusion
that the above-barrier cross sections of ${}^{46}$K$+{}^{181}$Ta are suppressed.

\begin{table}[h!]
\caption{ %
The barrier parameters calculated by the SPP and 
extracted from Ref.~\cite{Wakhle2018_PRC97-021602R}.
}
\begin{ruledtabular}
\begin{tabular}{c |ccc|cc}
 &             &      SPP      &  &  \multicolumn{2}{c}{Ref.~\cite{Wakhle2018_PRC97-021602R}}  \\ 
 \cline{2-6}\noalign{\vskip1pt}
Reaction& $V_{\rm B}$ & $\hbar\omega$ &  $R_{\rm B}$ & $V_{\rm B}$ &  $R_{\rm B}$ \\  
 & (MeV)       &    (MeV)      &  (fm)        & (MeV)       &  (fm) \\ \hline\noalign{\vskip3pt}
 ${}^{39}$K$+{}^{181}$Ta & 155.651  & 4.328 &  12.030 &  152.690  &  12.82 \\
 ${}^{46}$K$+{}^{181}$Ta & 152.141  & 4.029 & 12.333  &  146.484 &  10.16 \\ 
\end{tabular}\label{tab:para}
\end{ruledtabular}
\end{table}

The calculated capture cross sections of ${}^{39}$K$+{}^{181}$Ta 
and ${}^{46}$K$+{}^{181}$Ta are also reduced and shown in Fig.~\ref{fig:uff} by the 
dash and dash-dotted lines. It can be seen that, at sub-barrier energy region, the 
calculated cross sections are much larger than the UFF due to the coupled-channel effects.  
Furthermore, at sub-barrier energies, the reduced calculated capture cross sections 
of ${}^{39}$K$+{}^{181}$Ta are a little larger than those of ${}^{46}$K$+{}^{181}$Ta, 
which is owing to the coupling to the positive $Q$-value two-neutron transfer channel.
At energies above the Coulomb barrier, the predictions from the ECC model of these 
two reactions are consistent with the UFF. This means that, above the barrier, 
the measured capture cross sections of ${}^{46}$K$+{}^{181}$Ta are suppressed as 
compared with the predictions of the ECC model and the UFF. 
Therefore, for producing heavy and superheavy nuclei using the neutron-rich radioactive beams, 
it is necessary and important to consider this suppression.
Further experimental and theoretical studies are expected. 

\section{\label{sec:summary}Summary}
In summary, the capture cross sections for reactions ${}^{39}$K$+{}^{181}$Ta and ${}^{46}$K$+{}^{181}$Ta are 
investigated by using the empirical coupled-channel (ECC) model and the universal fusion
function (UFF) prescription. For the reaction ${}^{39}$K$+{}^{181}$Ta, 
the results of the ECC model are in good agreement with the data. While for the reaction 
with neutron-rich radioactive beam ${}^{46}$K, the predictions of the ECC model 
overestimate the above-barrier capture cross sections or, in other words, 
the measured capture cross sections are suppressed as compared with the ECC calculations. 
Comparing the reduced data of these two reactions, it is found that the data 
of above-barrier cross sections of ${}^{39}$K$+{}^{181}$Ta are consistent with 
the UFF and are larger than those of ${}^{46}$K$+{}^{181}$Ta. 
This implies that the capture cross sections of ${}^{46}$K$+{}^{181}$Ta are suppressed 
at energies above the Coulomb barrier. Furthermore, at sub-barrier energies, 
the reduced calculated capture cross sections of ${}^{39}$K$+{}^{181}$Ta 
are a little larger than those of ${}^{46}$K$+{}^{181}$Ta, which is owing to 
the coupling to the positive $Q$-value two-neutron transfer channel.

\acknowledgements
We thank Dr. A. Wakhle for providing us the data of the capture cross sections 
for reactions ${}^{39}$K$+{}^{181}$Ta and ${}^{46}$K$+{}^{181}$Ta and helpful discussions. 
The referee's suggestions on the expression of the empirical barrier weight function $f(B)$ 
are appreciated gratefully. 
This work has been partly supported by 
the National Key R\&D Program of China (No. 2018YFA0404402), 
the National Natural Science Foundation of China (Grants 
No. 11525524, 
No. 11621131001, 
No. 11647601, 
No. 11747601, 
No. 11711540016, and 
No. 11705165), 
the Key Research Program of Frontier Sciences, Chinese Academy of Sciences, 
the IAEA Coordinated Research Project ``F41033'', and 
the Physics Research and Development Program of Zhengzhou University (Grant No. 32410017).
The computational results presented in this work have been obtained on 
the High-performance Computing Cluster of KLTP/ITP-CAS and 
the ScGrid of the Supercomputing Center, Computer Network Information Center of 
the Chinese Academy of Sciences.

\end{document}